\documentstyle[aas2pp4]{article}          
\lefthead{ }
\righthead{ }
\begin{document}
\def\Msun{M_{\odot \hskip-5.2pt \bullet}}
\def\kms{km s$^{-1}$}
\def\r{\hangindent=1pc  \noindent}
\def\deg{$^\circ$}
\def\vrot{$V_{\rm rot}$}
\def\Vrot{V_{\rm rot}}
\def\ha{H${\alpha}$ }
\def\n2{$[{\rm{N}}_{\rm{II}}]$}
\def\h1{H${}_{\rm{I}}$}
\title{Iteration Method to Derive Exact Rotation Curves from
Position-Velocity Diagrams of Spiral Galaxies}
\author{Tsutomu TAKAMIYA and Yoshiaki SOFUE}
\affil{\it Institute of Astronomy, University of Tokyo, Mitaka,
Tokyo 181-0015, Japan\\
E-mail: sofue@ioa.s.u-tokyo.ac.jp }

\begin{abstract}
We present an iteration method to derive exact rotation curves (RC) of
spiral galaxies from observed position-velocity diagrams (PVD),
which comprises the following procedure.
An initial rotation curve, RC0, is adopted from an observed PV diagram (PV0),
obtained by any simple method such as the peak-intensity method.
Using this rotation curve and an observed radial distribution of
intensity (emissivity), we construct a simulated PV diagram (PV1).
The difference between a rotation curve obtained from this PV1
and the original RC (e.g., difference between peak-intensity velocities)
is used to correct the initial RC to obtain a corrected rotation curve, RC1.
This RC1 is used to calculated another PVD (PV2) using the observed
intensity distribution, and to obtain the second iterated RC (RC2).
This iteration is repeated until PV$i$ converges to PV0, so that
the differences between PV$i$ and PV0 becomes minimum.
Finally RC$i$ is adopted as the most reliable rotation curve.
We apply this method to some observed PVDs of nearby galaxies,
and show that the iteration successfully converges to give reliable rotation
curves.
We show that the method is powerful to detect central massive objects.
\end{abstract}

Subject headings: methods: data analysis ---
ISM: kinematics and dynamics ---
galaxies: ISM ---
galaxies: kinematics and dynamics ---
galaxies: nuclei ---
galaxies: structure ---

\section{Introduction}

Rotation curves (RC) are one of the most basic informations of dynamics of
galaxies (Sofue and Rubin 2001).
RCs are used to discuss supermassive black holes (e.g.,
Dressler \& Richstone 1988, Kormendy \& Richstone 1992, Bower et al. 1998),
the dark matter distribution in the halo
(e.g.,  Rubin et al. 1985, Kent
1986, Persic et al. 1996, Honma \& Sofue 1997, Takamiya \& Sofue 2000),
the characteristics of a bulge and disk (e.g., Kormendy \& Illingworth 1982,
Kent 1986, H\'{e}raudeau \& Simien 1997),
and kinematic peculiarities (e.g.,  M\'arquez \& Moles 1996,
Barton et al. 1999, Rubin et al. 1999).

RCs of disk galaxies are usually derived from position-velocity diagrams
(PVD) by optical (\ha, \n2) and radio (CO, \h1) line observations (SR2001).
There have been several ways to derive RCs.
Widely used methods are to trace intensity-weighted velocities
(Warner et al. 1973), and to trace peak-flux ridges in PVD
(e.g., Rubin et al. 1985, Mathewson et al. 1992).
These methods give good results for nearly face-on galaxies with sufficiently
high spatial resolutions.
Other methods are the terminal-velocity method used for our Galaxy
(e.g., Clemens 1985), and the envelope-tracing method
(Sofue 1996), which traces the envelope velocities on PVD and correct for
the intrinsic interstellar velocity dispersion and resolutions
to estimate terminal velocities.
This method gives better results for the central regions and for highly inclined and edge-on galaxies (Olling 1996).
Sofue (1996) applied this method also to intermediately inclined galaxies
to obtain central-to-outer RCs of many spiral galaxies.


The envelope-tracing method would be the most practical way to derive RCs
in disk galaxies.
However, it still has difficulties to derive exact velocities
in the central regions because of very high apparent velocity widths
due to unresolved rapidly-rotating components, whose radial and tangential
points are observed in a finite beam, due to steep velocity gradients, and
complex gas distribution.

In this paper, we first analyze the observational conditions
under which the peak-traced velocity does  not accurately follow the
true rotation velocity by simulating PV diagrams using model RCs.
Next, we propose a new iteration method to derive a RC, and apply
it to some observations and compare with the peak-flux and
envelope-tracing methods.
We finally stress the advantage to use the iteration method
to the search for central massive cores and black holes.

\section{PVD Simulation}

Given a RC, the shape of a PV diagram depends on the observational
parameters such as the seeing size, slit width, or equivalently the
beam size, and spectral resolution.
The PV shape also depends on the intrinsic parameters of
the galaxy itself such as the inclination of the disk, interstellar
velocity dispersion, and the gas distribution.
In order to see how these parameters affect the observed
PVDs, we simulate PVDs from a given RC for a model galaxy, and show the
result in figure 1.
Observational parameters such as the assumed slit width, velocity resolution
and seeing size are given on the top of the figure.

--- Fig. 1 ---

The model galaxy is put at a distance of 10 Mpc (1$''$=49pc).
The galaxy is assumed to have a similar RC to that of the Milky Way expressed
by a Miyamoto-Nagai potential model (Miyamoto \& Nagai 1975) with a
massive central core, bulge, disk and dark halo, as shown by the thick
line in the upper panel.
The gas disk has an exponential density profile
as indicated by the thin line in the lower panel, which is expressed by
\begin{equation}
\rho \propto \exp\left(-\frac{r}{{h}_{r}}-\frac{|z|}{{h}_{z}}\right),
\end{equation}
where $r$ $z$, $h_r$ and $h_z$ are the radius, height from the galactic plane,
the scale radius, and the scale height.
We assume that ${h}_{r}=1.5$(kpc) and ${h}_{z}=60$(pc), and the
inclination to be ${80}^{\circ}$.
The interstellar velocity dispersion of the order of 5 to 10 \kms
is assumed to be sufficiently small compared to the observational velocity
resolution, which is taken to be 35 \kms.

The middle panel of figure 1 shows the ratio of the 'observed'
peak-intensity velocity in the simulated PVD to the assumed rotation velocity.
The simulated PVD behaves rigid-body like in the central regions, and
hence, if we use peak-intensity velocities, the rotation velocity is
significantly underestimated.
We also made the simulation for various parameter sets, and found that
the larger is the inclination, the more is the underestimation, because we
observe more amount of foreground and background disk gases on the
line-of-sight with nearly zero radial velocities for higher inclination
galaxies.

Moreover, if the central region is gas deficient, namely if the gas
distribution is ring like, the observed PV diagram behaves more
rigid-body like, even if the assumed central rotation velocity was
extremely high.
Underestimation of central rotation velocities was found in all of cases,
and it amounted to 50 \% to 100 \% of the intrinsic rotation velocities.
The underestimation were found even for the envelope-tracing method,
although it gave better results than the peak-tracing method.
Hence, the current widely used methods may not be appropriate to discuss
rotation curves and related properties such as the mass distribution
in the central regions of spiral galaxies.

\section{Iteration Method}

In order to derive more reliable RCs from observed PV diagrams, particularly
for the central regions of spiral galaxies, we propose
a new method, which comprises the following algorithm,
which we call hereafter the iteration method.
Figure 2 shows the flow-chart of this algorithm.

--- Fig. 2 ---

(1) We define a radial velocity profile traced at 20 \%-level envelope
    of the peak flux in a PV diagram as a comparison velocity, and
    take this profile as the initial trial RC, $V_{\rm ini}(r)$.

(2) At each radius we calculate velocity-integrated intensity using the
    observed spectrum, or equivalently by integrating the PV diagram
    in the direction of velocity at a fixed radius.
    We assume that the integrated intensity is proportional to the column
    density of interstellar gas along the line of sight, $\Sigma(r)$.
    The gas density distribution, $\rho(r, z)$, in the galaxy is assumed to
    have a disk form with an exponential $z$ directional structure:
\begin{equation}
\rho(r,z) = A \Sigma(r)\times \exp\left(-\frac{|z|}{h_z}\right).
\end{equation}
    Here, $h_z$ is the scale height of the disk, and assumed to be constant
    at 60 pc for all galaxies.
    The constant coefficient $A$ is taken to be arbitrary, because the
    absolute values of PV intensities do not affect the resultant rotation
    velocities in the present method.
    The thus once calculated $rho(r,z)$ is used through the entire
    iteration process.

(3) Based on $V_{\rm ini}(r)$ and $\rho(r,z)$, a PV-diagram is calculated 
    using
 observational parameters such as the slit width, velocity 
    resolution, and seeing size (beam size), which are taken to mimic the 
    real observations.
    We use this new PV diagram to derive a new 20 \%-level envelope of the
    peak flux, and obtain a new RC with velocities $V_{\rm com}$.

(4) We define the difference between the first trial RC and this calculated
    RC by $\delta V=V_{\rm ini} - V_{\rm com}$, and use it to correct
    for $V_{\rm ini}$ to obtain the second iterated RC,
    $V_2=V_{\rm ini}+\delta V$.

(5) We calculate another PV diagram using $V_2$ and obtain the second
    iterated RC, $V_{\rm com,2}$,
    which is then compared with $V_{ini}$ to calculate
    $\delta V_2=V_{\rm ini}-V_{\rm com,2}$, and obtain $V_3=V_2+\delta V_2$.

(6) We repeat this procedure for $i$ times to obtain the $i$-th iterated RC,
    $V_i=V_{i-1}+\delta V_i$, until $\delta V_i$ becomes smaller than a
    criterion, e.g. until $|\delta V_i|$ becomes sufficiently small compared to
    the velocity resolution.
    Here, $V_{{\rm com}, i}$ converges to $V_{\rm ini}$ within the error,
    and the calculated RC becomes approximately identical to the observed
    RC within the rms noise.
    Finally, we adopt $V_i$ as the most reliable rotation curve.

Here, we used the original $\Sigma$ obtained from observation to
calculate the iterated PVDs, whereas we used corrected velocities.
In order to obtain fully consistent iteration, the surface density must
also be replaced by corrected ones.
However, this would make the program very sophysticated, and will
be a subject for the future.
We only mention that the correction for $\Sigma$ would be much less
effective compared to the correction to velocities, because the
gas distribution is not expected to change so drastically compared to
velocities, which may have an extremly steep rise near the nucleus.

\section{Application to Observational Data}

We applied the iteration method to optical spectral data obtained by using
the 1.88-m telescope at Okayama Astrophysical Observatory (Sofue et al. 1998),
and to the CO ($J=1-0$) line data observed with the Nobeyama mm-wave Array
(NMA) (Sofue et al. 2001; Koda et al. 2002; Sofue et al. in preparation).
Here, we display some examples of the results applied to NMA
CO-line observations.

Figure 3a (left panel) shows an original PVD for the edge-on Sc galaxy
NGC 3079 in the CO line emission, exhibiting a central high-velocity
rotating molecular disk.
Figure 3b (middle) shows the obtained rotation curve by applying the
iteration method, and a constructed PVD by conniving this RC with
the observed intensity profile.
In figure 3c (right), we show a simple rotation curve obtained by the
peak-tracing method, which corresponds to the initial trial RC in figure 2,
and a convoluted PVD.
The iteration method gives an extremely steep rotation curve, and reproduces
the observed PVD very well.
On the other hand, the peak-tracing method gives a mild rotation curve,
but the reconstructed PVD cannot reproduce the observation.

Figures 3d to 3f are the same, but applied to a mildly inclined Sb galaxy
NGC 4536.
Again, we find the iteration method gives a reasonable reproduction of the
observed PVD, while the other method cannot reproduce the observation.
In both cases, peak-intensity velocities lead to underestimated rotation
velocities by 50 to 100 \kms in the central regions.
In figure 4 we show some more examples of rotation curves obtained by the
iteration method superposed on the original PVDs.
Very steeply rising rotation curves are obtained for most cases.

--- Fig. 3, 4 ---

\section{Discussion}

The iteration method has some advantage to the current methods, and
will be particularly powerful to determine the central rotation curves,
and therefore, to detect massive central objects.
In fact most of the galaxies shown in figures 3 and 4 have
very high velocities near the centers, which suggest the existence of
massive cores around the nuclei.
Underestimation of RCs in the central region significantly
affects the discussion of the central structure (Takamiya and Sofue 2000).

The iteration method uses all data points to reproduce the observed
PVDs by simulating the observed PVD, and hence, the statistical error
in the results are smaller compared to other methods.
For example, the peak-tracing and envelope-tracing methods uses only
a part of each velocity profile.
Hence, the amount of data available to fit observation is by a factor of
ten greater in the present method than the current method.
The intensity-weighted velocity method uses all data, but smears out the
detailed velocity information, and the result is approximately the same as
that from the peak-tracing method, which also significantly underestimate
the central velocities.

Finally, we stress that the present method does not need any potential
model, and therefore, the result is purely observational and unique.
This unique rotation curve can be further used to calculate the mass
distribution directly by a deconvolution technique as described in
Takamiya and Sofue (2000).
A method comparing the shapes of observed and calculated PV diagrams
has been used by Bertola et al. (1998) to detect central massive objects.
They assumed a central potential model in order to mimic the PV diagrams, and
hence, the method cannot measure the error quantitatively, and
the mass model may not necessarily be unique.
Application of the iteration technique to these current observations would
also give more reliable answer to the existence of such massive objects.

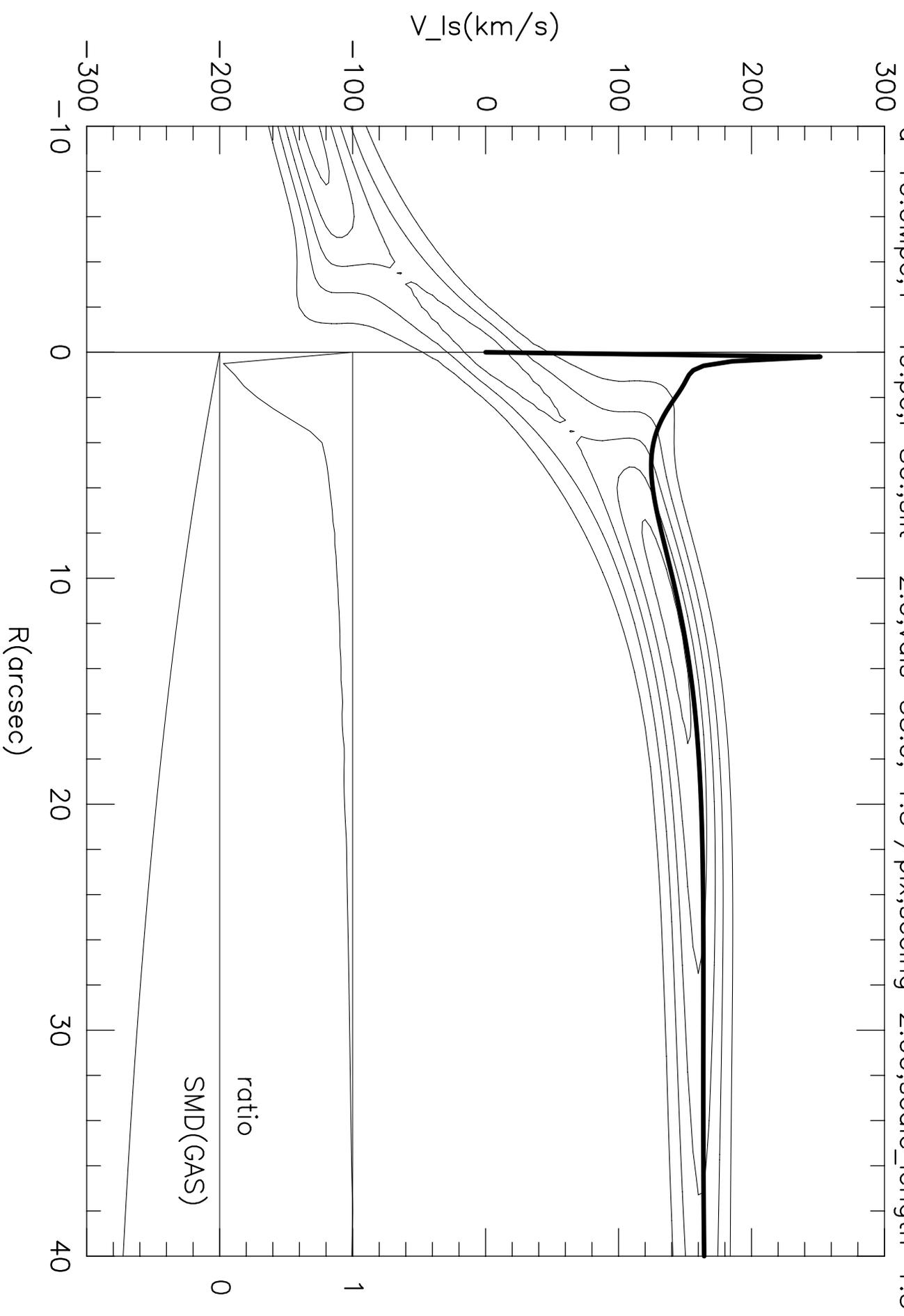
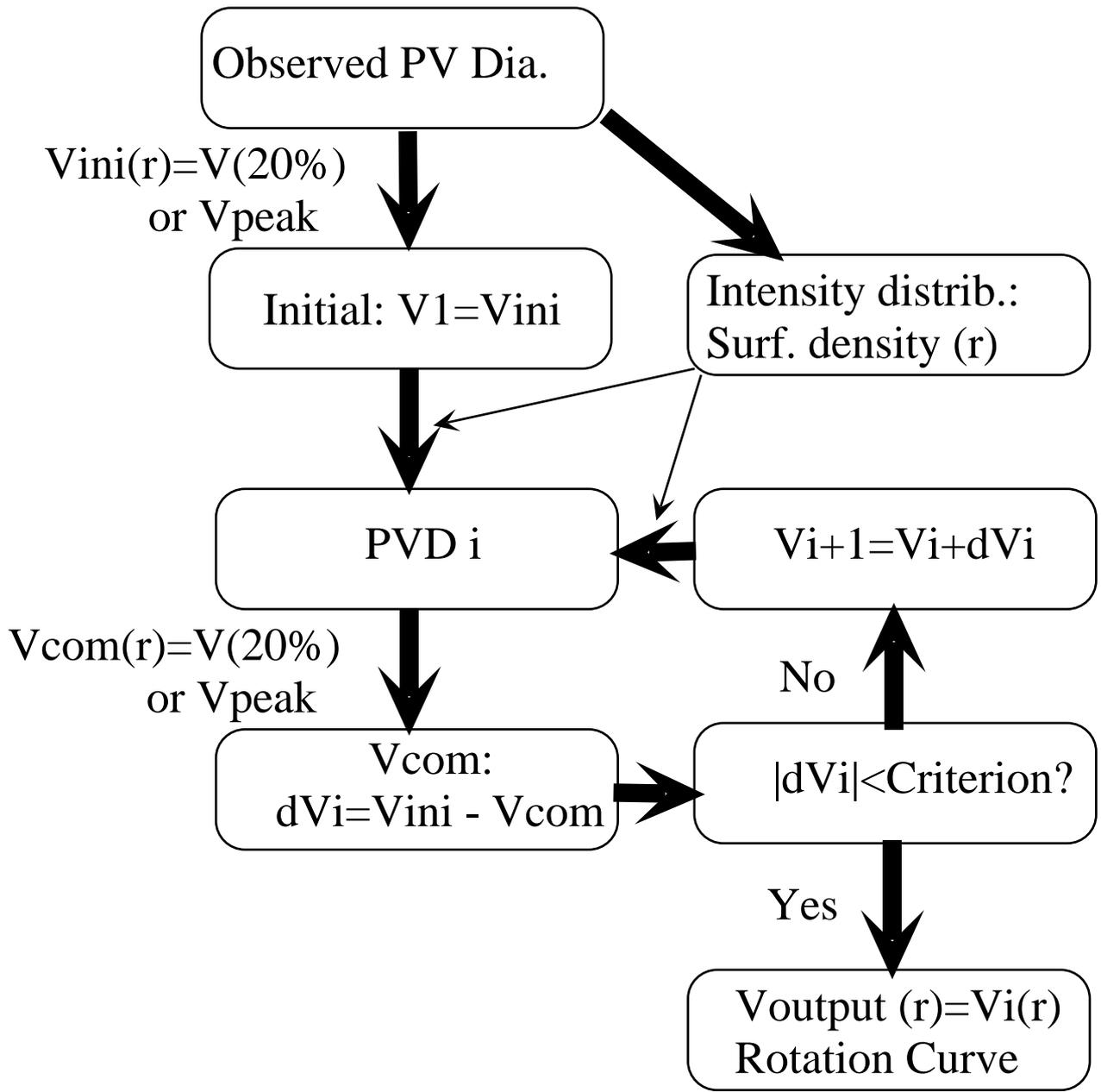
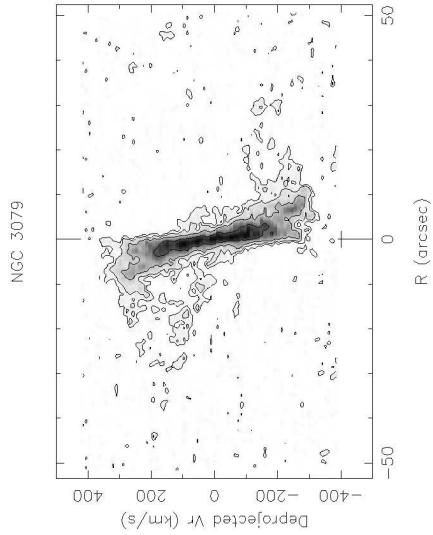
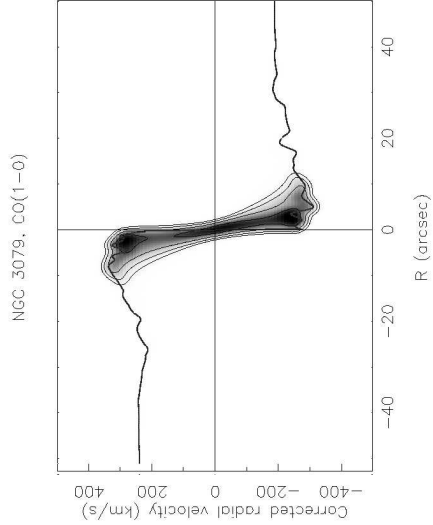
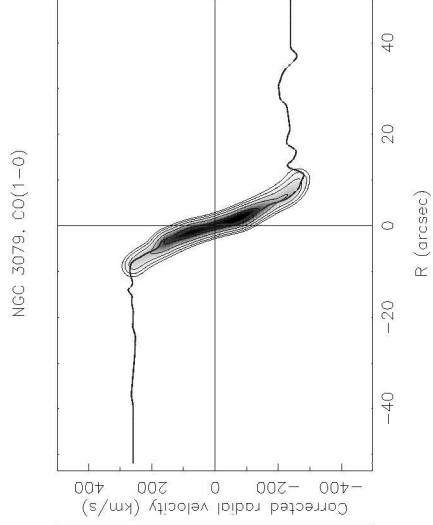
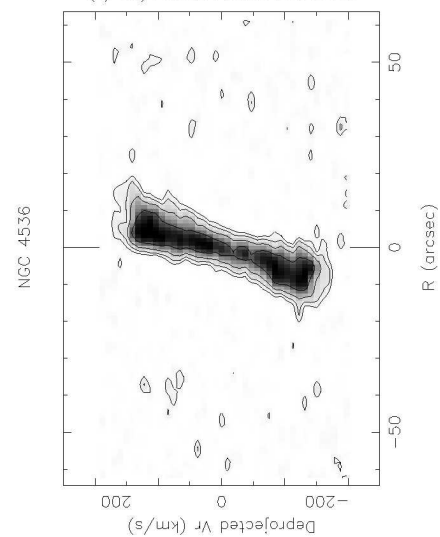
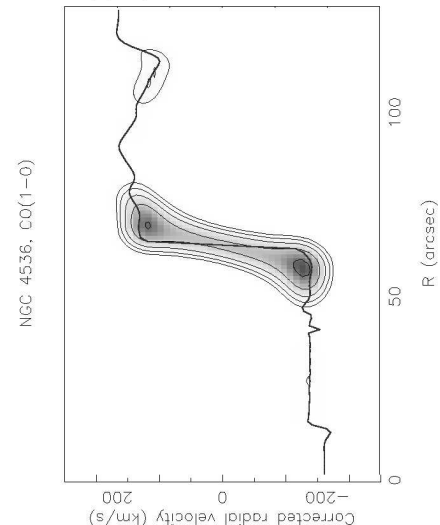
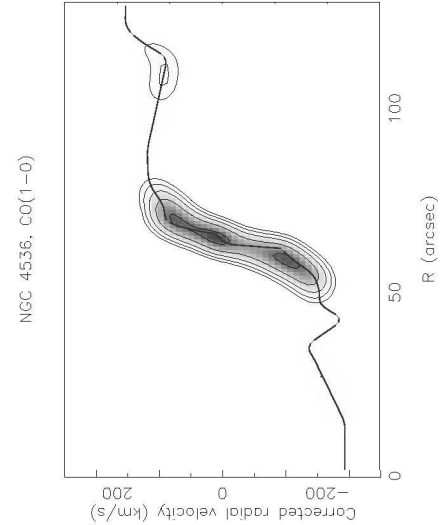
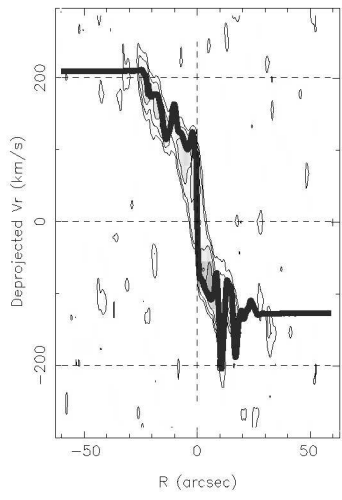
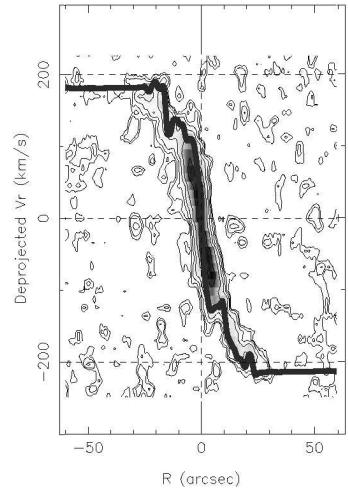
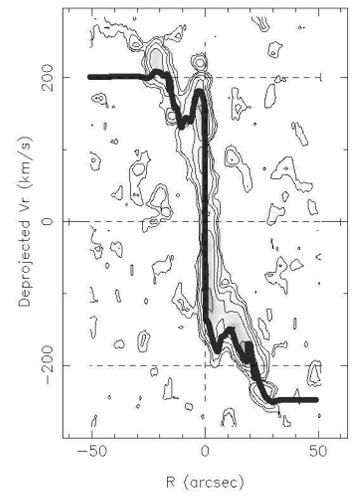
\end{document}